\documentclass[1p]{Article_website}
\usepackage{graphicx}
\usepackage{amsmath}
\usepackage[misc,geometry]{ifsym}
\usepackage{enumitem}
\newlist{symb}{itemize}{1}
\setlist[symb,1]{label=,labelwidth=15mm,align=parleft,itemsep=0.1\baselineskip,leftmargin=!}
\usepackage{color}

\title{Bifurcation Dodge: Avoidance of a Thermoacoustic Instability under Transient Operation}

\author{Giacomo Bonciolini, Nicolas Noiray}

\address{CAPS Laboratory, MAVT department ETH Z\"urich, Sonneggstrasse 3, 8092, Zurich, Switzerland}
\begin{document}

\begin{abstract}
    Varying one of the governing parameters of a dynamical system may lead to a  critical transition, where the new stable state is undesirable. In some cases, there is only a limited range of the bifurcation parameter that corresponds to that unwanted attractor, while the system runs problem-less otherwise. In this study, we present experimental results regarding a thermoacoustic system subject to two consecutive and mirrored supercritical Hopf bifurcations: the system exhibits high amplitude thermoacoustic limit cycles for intermediate values of the bifurcation parameter. Changing quickly enough the bifurcation parameter, it was possible to dodge the unwanted limit cycles. A low-order model of the complex thermoacoustic system was developed, in order to describe this interesting transient dynamics. It was afterward used to 
    assess the risk of  exceeding an oscillation amplitude threshold as a function of the rate of change of the bifurcation parameter.      
\end{abstract}
\begin{keyword}
Hopf Bifurcation \sep Transient \sep Escape \sep Thermoacoustic \sep Parameter Identification\\
\vspace{5mm}
\textcopyright2018. This manuscript version is made available under the CC-BY-NC-ND 4.0 license http://creativecommons.org/licenses/by-nc-nd/4.0/ 
\end{keyword}

\maketitle
\twocolumn
\section*{List of symbols}
\begin{symb}
\item[$t$]{Time [s]}
\item[$T_\text{r}$]{Ramp duration [s]}
\item[$R$]{Ramp rate}
\item[$T_\nu$]{Characteristic oscillation growth time [s]} 
\item[$\dot{m}_\text{air}$]{Air mass flow rate [kg/s]}
\item[$\dot{m}_\text{ng}$]{Natural Gas mass flow rate [kg/s]}
\item[$\phi$]{Equivalence ratio}
\item[$p$]{Acoustic pressure [bar]}
\item[$A$]{Acoustic oscillation random amplitude [bar]}
\item[$\varphi$]{Oscillation random phase [rad]}
\item[$\omega$]{Angular frequency [rad/s]}
\item[$S_{pp}$]{Power Spectral Density of the signal $p$ [$[p]^2$ Hz$^{-1}$]}
\item[$P(X)$]{Probability Density Function (PDF) of a stochastic variable $X$ $[X]^{-1}$}
\item[$P_\text{max}$]{=$\text{max}_X\{P(X)\}$}
\item[$\omega_0$]{Oscillator natural angular frequency [rad/s]}
\item[$\nu$]{Linear growth rate [rad/s]}
\item[$\nu_\text{min}$]{Minimum $\nu$ of the range [rad/s]}
\item[$\nu_\text{max}$]{Maximum $\nu$ of the range [rad/s]}
\item[$\kappa$]{Cubic saturation constant [rad/s $[p]^{-2}$]}
\item[$\xi(t)$]{White noise process [$[p]$/s$^2$]}
\item[$\Gamma$]{Noise intensity}
\item[$A_\text{th}$]{Threshold Amplitude}
\item[$\langle A(t) \rangle$]{Expected value of $A(t)$}
\item[$A_\infty(\dot{m}_\mathrm{air})$]{$=\int_0^\infty AP_\infty(A;\,\dot{m}_\mathrm{air})\text{d}A$}
\item[$A_\text{min}$]{=$\text{min}\{A_\infty\}$ (when $\nu=\nu_\text{min}$)}
\item[$A_\text{max}$]{=$\text{max}\{A_\infty\}$ (when $\nu=\nu_\text{max}$)}
\item[$\Pr_\text{nc}(t)$]{$=\int_0^{A_\text{th}}P(A;\,t)\text{d}A$}
\item[$\Pr_\text{d}$]{=$\Pr_\text{nc}(T_\text{r})$ Dodge Probability }
\item[$\Delta t_\text{th}$]{Time interval spent over the threshold $A_\text{th}$}
\item[$E$]{=$\int_{\Delta t_\text{th}}{\langle A(t) \rangle}^2\text{d}t$ Mean energy}
\end{symb}
\onecolumn
\section{Introduction}
\begin{figure*}[th!]
    \centering
    \includegraphics[width=\textwidth]{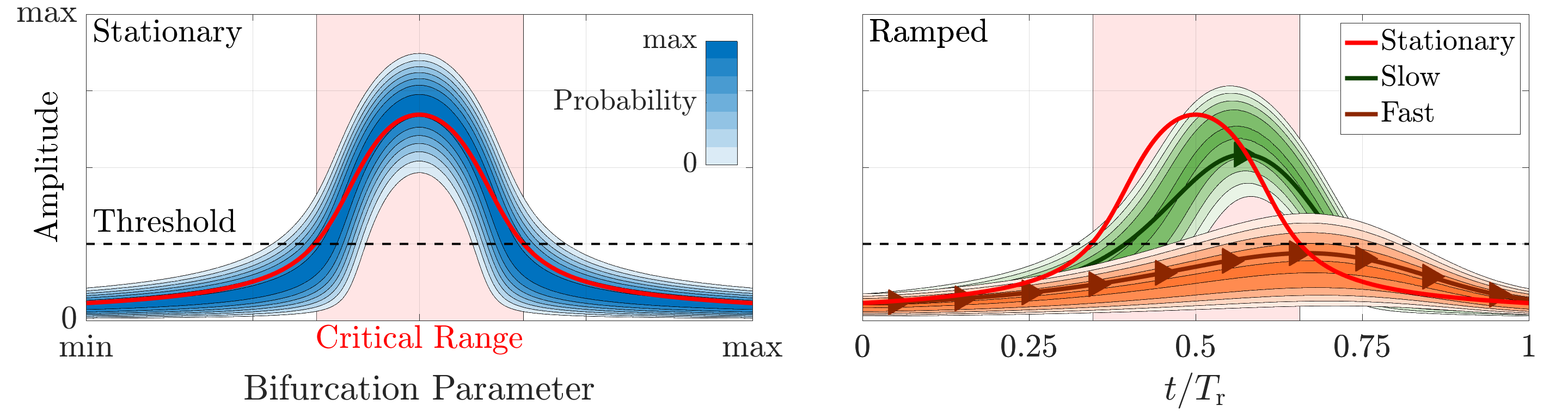}
    \caption{Context of the paper. Left:  Stochastic bifurcation diagram of two consecutive and symmetric supercritical bifurcations. Given a threshold for the amplitude, a critical  range for the bifurcation parameter is defined. This diagram is valid for quasi-steady variation of the bifurcation parameter. Right: possibility of dodging the high-amplitude, by ramping quickly enough the bifurcation parameter.}
    \label{fig:01_intro}
\end{figure*}
%
Many complex dynamical systems exhibit critical transitions that are characterized by a significant change of their behaviour when one of their governing parameters exceeds a certain value. This type of tipping points results from bifurcations of quasi-steady attractors \cite{ashwin2012}: for a small change of one of its parameters, the topology of the dynamical system phase portrait markedly changes \cite{kuznetsov2013elements}. Among countless examples, one can refer to   \cite{hoegh2007coral} where the coral reef extinction is associated to a change in atmospheric $\text{CO}_2$ concentration, to \cite{bosi2018limit} where the increase of the environment pollution leads to the emergence of consumption demands limit cycles, to \cite{lenton2008tipping} where the influence of different parameters on  climate and on ecosystems catastrophes is reviewed, or to \cite{kuehnen2015subcritical} where a supercritical Hopf bifurcation of the flow in a helical pipe is observed when the Reynolds number exceeds a certain value. In all these examples, the system bifurcates to undesirable and/or dangerous states, which can be extremely detrimental for the integrity of the system. This is why the development of methods to detect early-warning signals or precursors of a bifurcation is an active research area  \cite{scheffer2012anticipating,kuehn2011mathematical,karnatak2017early,dakos2015resilience,drake2010early,ritchie2016early,lenton2012early,nikolaou2015detection,chen2018forecasting,yang2018stochastic,kim2018predicting}.\\ 
In most cases, the system displays an unwanted behavior for a limited interval of the bifurcation parameter, while it runs problem-free otherwise (see for instance \cite{bilinsky2018slow}). This type of situation is treated in the present study, with stochastic forcing of the deterministic system as an additional ingredient. 
One can refer to recent examples of stochastic bifurcations in biology \cite{chamgoue2013bifurcations,cao2018linear,guo2018stochastic}, psychology \cite{yue2015stochastic}, medicine and epidemiology \cite{xu2013stochastic,haldar2015bifurcation}, economics \cite{chiarella2008stochastic,zhang2016dynamical}, chemistry \cite{bishop2010stochastic} or oscillators mechanics \cite{yilmaz2018stochastic,zakharova2010stochastic,wang2018quasi,yang2018bifurcation,huang2018analysis,venkatramani2018intermittency,kumar2016stochastic}.
The stochastic bifurcation considered here is depicted in fig. \ref{fig:01_intro}a. When the bifurcation parameter takes values within the critical range, which is in the middle of the considered interval, the mean of the output amplitude of the dynamic system exceeds the \textit{acceptable} threshold. Due to the stochastic forcing, the amplitude is described in terms of its probability density function (PDF). The red line represents the mean of the distribution for each value of the bifurcation parameter. This dynamic system has a non monotonic behaviour: it is characterized by small amplitudes in the lower and upper range of the bifurcation parameter, while it leads to high amplitudes for intermediate values.\\
The aim of the present study is to investigate at which rate the bifurcation parameter must be ramped through the critical range, in order to dodge the high amplitudes. The amplitude statistics of two such transients are illustrated in fig. \ref{fig:01_intro}b. Note that the two ramp durations $T_\text{r}$ are different and that the time axis in this diagram has been normalized. While the slow ramp is ineffective in avoiding the transition to high amplitudes (green contour), a fast ramp (orange contour), due to inertial effects,  enables a bifurcation dodge, in the sense that the ensemble average of the amplitude for a large number of ramp realization remains below the threshold.\\
In the following section, the system used to illustrate  such bifurcation dodge and its low order modeling is presented. It consists of a thermoacoustic instability in a laboratory scale burner. This is an unwanted phenomenon in practical combustors because the resulting intense acoustic field induces harmful vibrations. For example, aero-engine combustors are designed such that no thermoacoutic instabilities exist for cruise and take-off conditions. This task is a true challenge for today and tomorrow's engines, because combustor optimization for reduction of pollutant emissions is very often accompanied with a narrowing of the  ``thermoacoustics-free'' operating window. It therefore constitutes a good example of systems for which a theoretical framework for assessing the possibility dodging bifurcations is highly relevant.
\section{Thermoacoustic instabilities}
\begin{figure}
    \centering
    \includegraphics[trim=200 50 0 0,width=0.7\columnwidth]{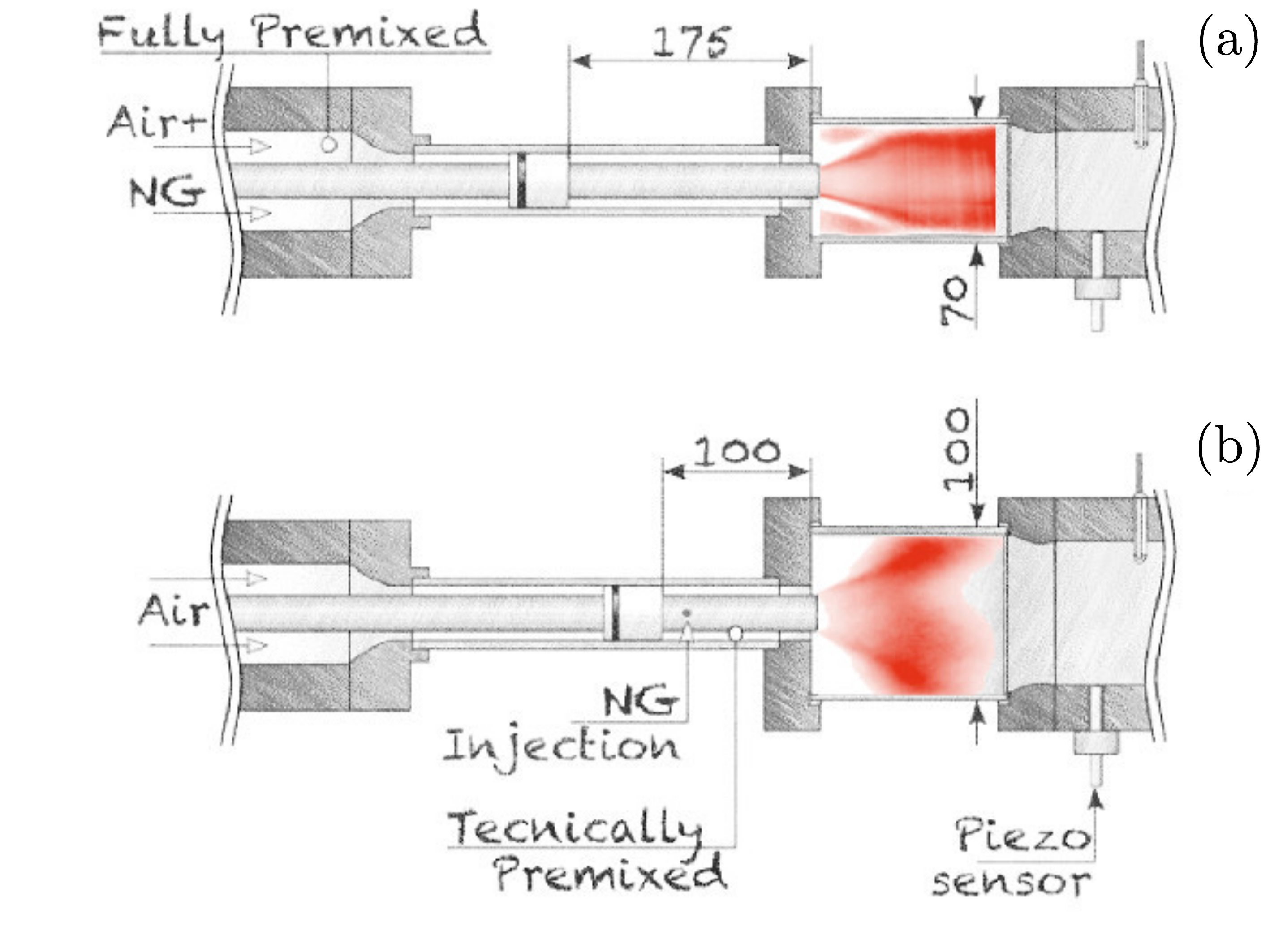}
    \caption{Comparison between (a) the geometry of the combustor configuration used in \cite{bonciolini2018effect}, and (b) the one used in the present study.}
    \label{fig:02_geometry}
\end{figure}

Thermoacoustic instabilities are a challenging problem encountered in many practical combustion systems, such as heavy-duty gas turbines, aeronautical or rocket engines \cite{poinsot2017prediction}. This phenomenon arises due to the coupling between unsteady heat release rate and the acoustic field in the combustor \cite{lieuwen2012book}. Under certain conditions, this interaction becomes constructive, leading to extremely high acoustic pressure oscillations. This acoustic field loads the mechanical components and can lead to fatigue failures. Therefore, the thermoacoustic stability of a new combustor must be assessed in the design phase, to avoid any unstable operation. \\
The linear stability depends on the flame response to acoustic perturbations, which depends on the operating condition. Often, the thermoacoustic stability does not vary monotonically with changes of operating parameters, although the mean flame shape is not significantly altered. In these cases, the non-monotonic stability behaviour is associated with a change of the phase lag between the flame response to acoustically-triggered flow perturbations and the local acoustic field \cite{ghirardo2018effect}. 
In other cases, a change of the operating parameters leads to a substantial modification of the flame topology, which can be accompanied with a thermoacoustic instability \cite{Shanbhogue2016,bonciolini2018effect}. In \cite{bonciolini2018effect}, it was shown that when the thermal power is increased by ramping the fuel mass flow up, the rise of the temperature of the combustion chamber walls  strongly impacts the flame shape and consequently the thermoacoustic stability. In that particular configuration, and in contrast with the present study, it was demonstrated that it is impossible to dodge the limit cycles with a fast ramping of the fuel mass flow, because of the wall thermal inertia and of the significant effect of the wall temperature on the mean flame shape.\\
In the present study, the lab-scale combustor has been modified, as shown in fig. \ref{fig:02_geometry}. The main differences between the combustor configuration (a) used in  \cite{bonciolini2018effect} and the present configuration (b) are the following:
\begin{itemize}
    \item The combustion chamber diameter was increased from 70 to 100 mm, which sets a lower confinement of the flame. Consequently, the flame shape is less sensitive to changes of the wall temperature for the considered range of fuel and air mass flows.
    \item The swirler was moved 75mm closer to the lance tip. The latter defines the inner hot combustion products recirculation zone, which governs the anchoring of the swirling flame.
    \item The combustor is not operated in fully premixed mode, but in technically premixed mode, with natural gas injection downstream of the swirler.
\end{itemize}
More information about this lab-scale combustor can be found  in \cite{bonciolini2018effect}.
Its thermoacoustic behavior under stationary operating conditions, where fuel and air mass flows are kept constant, is presented in the following section.
%

\section{Stationary operation}
\begin{figure}
    \centering
    \includegraphics[width=0.5\columnwidth]{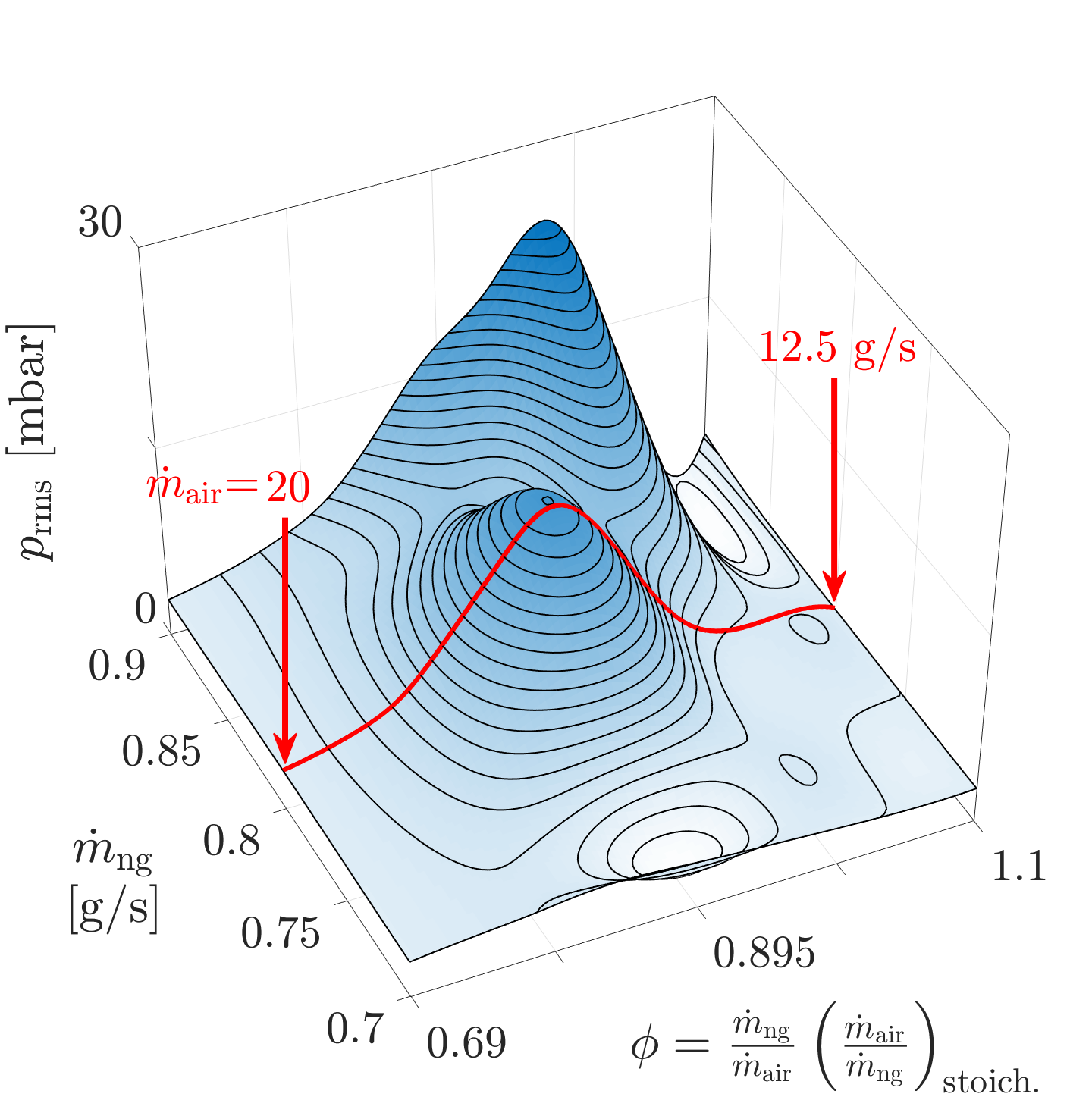}
    \caption{Root mean square of the acoustic pressure in the combustion chamber, as a function of the operating condition, defined by the natural gas mass flow rate $\dot{m}_\text{ng}$ and the equivalence ratio $\phi$. The red line is the path that is analyzed in the remainder of this study.}
    \label{fig:03_map}
\end{figure}

The combustor presented in fig. \ref{fig:02_geometry}b can be operated  at different conditions, by adjusting the air and natural gas (NG) mass flows $\dot{m}_\text{air}$ and $\dot{m}_\text{ng}$. It exhibits thermoacoustic instabilities for a large portion of the operating conditions. The acoustic pressure was recorded by a water-cooled Kistler piezoelectric sensor type 211B2  placed downstream of the flame in the combustion chamber (see fig. \ref{fig:02_geometry}b). Fig. \ref{fig:03_map} shows a map of the root mean square of the acoustic pressure $p_\text{rms}$ as a function of the two operating parameters that were varied to characterize the thermoacoustic behaviour of the system: the equivalence ratio $\phi=(\dot{m}_\text{ng}/\dot{m}_\text{air})/(\dot{m}_\text{ng}/\dot{m}_\text{air})_\text{stoichiometric}$ and the natural gas mass flow rate $\dot{m}_\text{ng}$. The explored range was $\phi\in$ [0.69; 1.1] and $\dot{m}_\text{ng}\in$ [0.7; 0.9] g/s, corresponding to a thermal power ranging from  26.6 to 34.2 kW. From this map, one can notice that the system exhibits high amplitude limit cycles in two regions, one located at intermediate $\dot{m}_\text{ng}$ and $\phi$ and the other at high $\dot{m}_\text{ng}$ and $\phi$.\\
\subsection{Thermoacoustic bifurcation}
\begin{figure*}
    \centering
    \includegraphics[width=\textwidth]{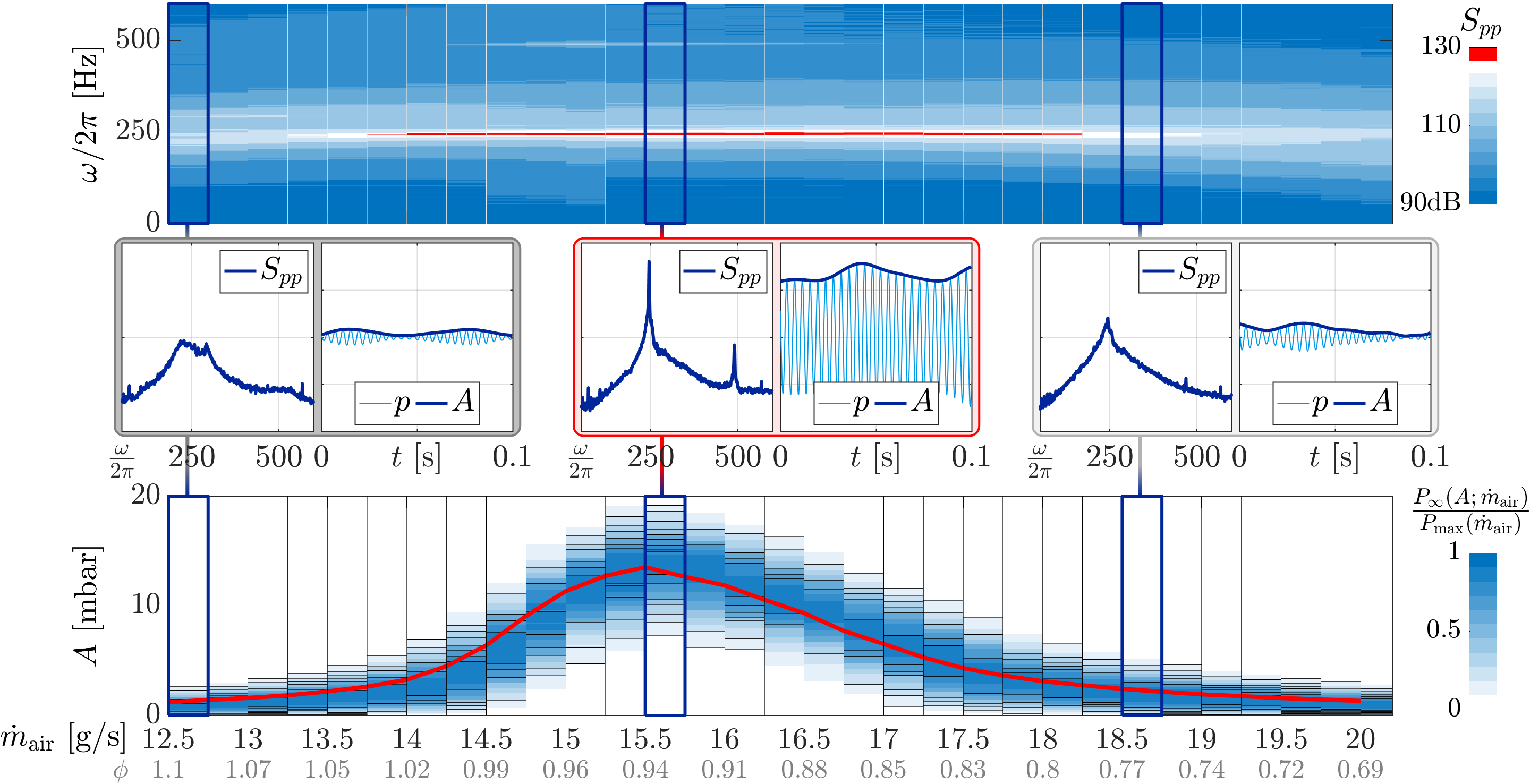}
    \caption{Stationary operations of the combustor at $\dot{m}_\text{ng}$=0.8 g/s, as a function of $\dot{m}_\text{air}$. Top: Acoustic pressure power spectral density.
    Bottom: Acoustic pressure oscillation amplitude probability density function, normalized with the maximum of the distribution at each operating condition. Center: details of spectra and time traces of three selected operating conditions: $\dot{m}_\text{air}$= 12.5, 15.5, 18.5 g/s.}
    \label{fig:04_bifurcation}
\end{figure*}
\begin{figure}
    \centering
    \includegraphics[width=0.49\columnwidth]{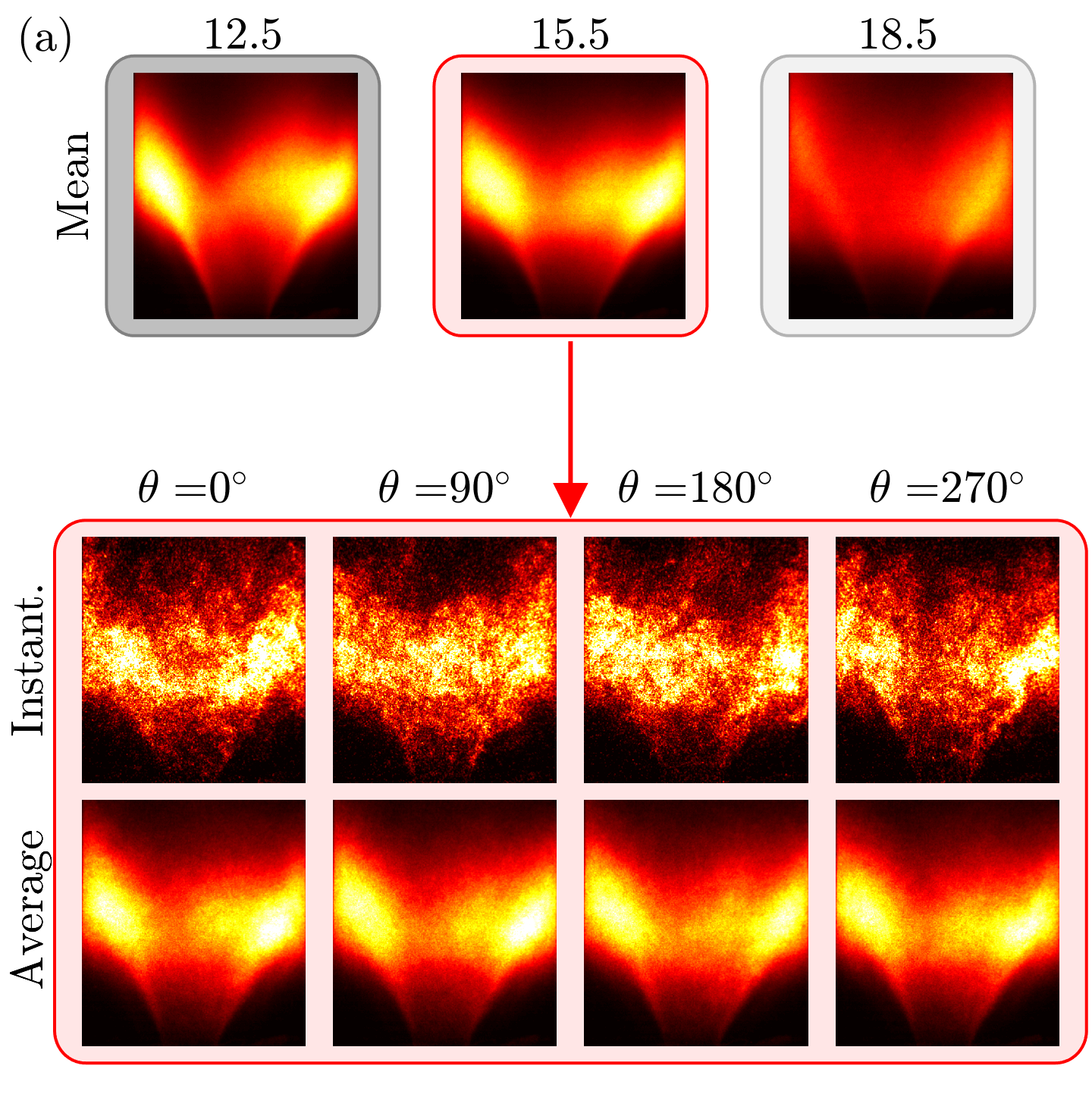}
    \includegraphics[width=0.49\columnwidth]{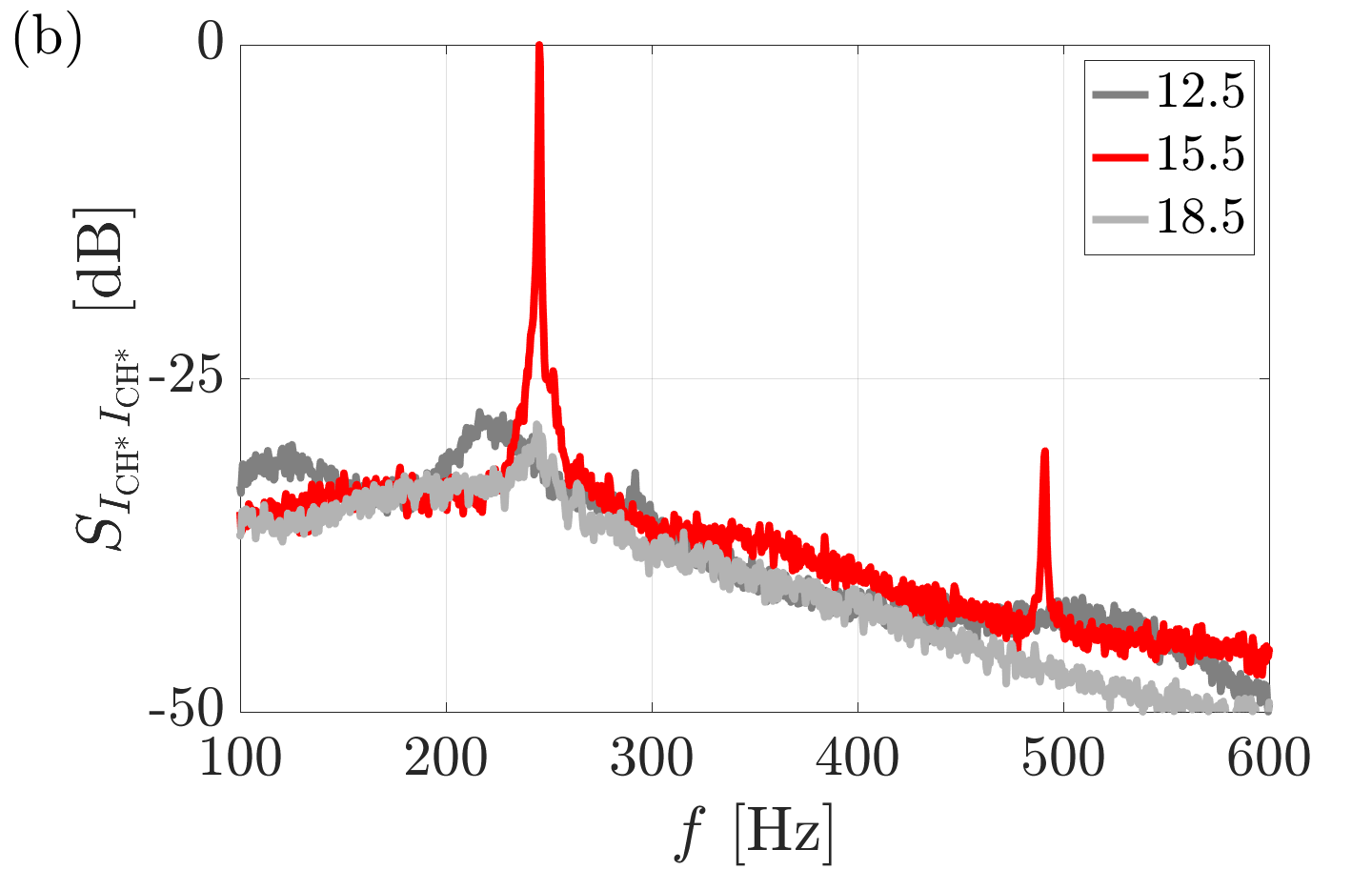}
    \caption{a) Top: High-speed camera acquisitions of the flame mean CH* chemiluminescence at three different operating conditions. From left to right: $\dot{m}_\text{air}$= 12.5; 15.5; 18.5 g/s.  Bottom: Detail of the $\dot{m}_\text{air}$= 15.5  g/s case, with instantaneous snapshots and corresponding phase-averaged CH* chemiluminescence at four different phases $\theta$ of the acoustic cycle $p(t) = A(t)\cos(\theta(t))$. b) Power spectral density of the CH$^*$ chemiluminescence integrated signal.}
    \label{fig:05_flames}
\end{figure}

In the remainder of this study we will focus on the bifurcation path that is highlighted with a red line on the map of fig. \ref{fig:03_map}; the combustor was operated with a constant $\dot{m}_\text{ng}$ of 0.8 g/s, varying the air mass flow rate $\dot{m}_\text{air}$ -which serves as bifurcation parameter- from 12.5 to 20 g/s. In that way, the whole equivalence ratio range was covered.\\
The detailed analysis of this bifurcation is shown in fig. \ref{fig:04_bifurcation}. The data presented here are obtained post-processing 2 minutes long acoustic pressure $p(t)$ acquisitions at stationary condition. In the upper panel, the power spectral density (PSD)  $S_{pp}$ at each operating condition is plotted. One can notice how a mode at circa 250 Hz is present at any condition of the examined range. In the central part of the range, the quality factor of the 250 Hz peak increases, and at the same time a peak at the first harmonic appears. These facts indicate the emergence of a thermoacoustic limit cycle. In the present case, where one mode dominates the thermoacoustic dynamics, the acoustic pressure in the combustion chamber can be approximated as $p(t,\boldsymbol{x})\approx A(t)\cos(\omega_0t+\varphi(t))\psi_0(\boldsymbol{x})$, where $\omega_0$ is the natural angular eigenfrequency of the eigenmode $\psi_0$, and $A(t)$ and $\varphi(t)$ two slowly-varying quantities representing the amplitude and phase fluctuations. These two quantities fluctuate because this thermoacoustic system is subject to a stochastic forcing exerted by turbulence, which relentlessly drives the system away from its deterministic attractor. The oscillation amplitude $A(t)$ is therefore a random variable with stationary distribution $P(A)$. In the lower panel $P(A)$ is presented (contour map) together with the distribution mean (red line) as a function of $\dot{m}_\text{air}$. This representation confirms what was already deducible from the power spectral density plot. The system undergoes two consecutive and mirrored supercritical Hopf bifurcations. At low $\dot{m}_\text{air}$, the system is linearly stable and the recorded fluctuations are only due to the turbulence stochastic forcing. At intermediate air mass flow rates, the system loses its linear stability: a stochastically perturbed limit cycle is established. For high $\dot{m}_\text{air}$, the system is linearly stable again, with small amplitude oscillations that are characteristic of randomly forced weakly damped oscillator. These three steps are exemplified by the three operating points $\dot{m}_\text{air}$= 12.5 g/s (stable), 15.5 g/s (unstable), 18.5 g/s (stable) in the central row of fig.  \ref{fig:04_bifurcation}, where PSDs and portions of the acoustic time traces are plotted.\\
This behaviour is analogous to the one presented in \cite{bonciolini2018effect}. In the present work, however, the instability is not accompanied by a flame topology transition (from V-shape to M-shape in  \cite{bonciolini2018effect}), and remains as a V-shape flame for the entire range of $\dot{m}_\text{air}$ as displayed in fig.~\ref{fig:05_flames}a. These high speed images were recorded at a frame rate of 5 kHz with a CMOS camera (Photron FASTCAM SA-X2) coupled to a high-speed intensifier (LaVision HS-IRO) equipped with a 100 mm f/2.8 UV lens (Cerco, aperture f\#6) and a bandpass filter for CH* (Edmund, transmission $>$90\% at 430 nm, FWHM 10 nm).
In the top row, the mean flame shape at the three operating conditions $\dot{m}_\text{air}$= 12.5, 15.5, 18.5 g/s is shown. For the unstable case $\dot{m}_\text{air}$= 15.5 g/s, phase-averaged information is given in the bottom panel. Each column represents one phase $\theta$ of the acoustic cycle $p(t)=A(t)\cos(\theta(t))$, and it is split in two rows; the top and bottom rows respectively show instantaneous snapshots and phase-avareged images of the flame CH* chemiluminescence. One can note that the coherent motion of the flame is not easy to perceive from these images, although the  coherent component of the heat release rate $\dot{Q}(t)=\int_V\dot{q}(t,\boldsymbol{x})\text{d}\boldsymbol{x}$ enables the high amplitude self-sustained thermoacoustic oscillations. In that regard, it is interesting to show the power spectral density of the chemiluminescence measurement $S_{I_\text{CH*}I_\text{CH*}}$ (see fig.~\ref{fig:05_flames}b), albeit $\dot{Q}$ is not completely proportional to the flame chemiluminescence $I_\text{CH*}$ in this technically premixed configuration \cite{najm1998adequacy,lauer2011determination,boujo2016quantifying}.

%
\subsection{Model and Identification of parameters}\label{ss:vdp}
\begin{figure}
    \centering
    \includegraphics[width=0.5\columnwidth]{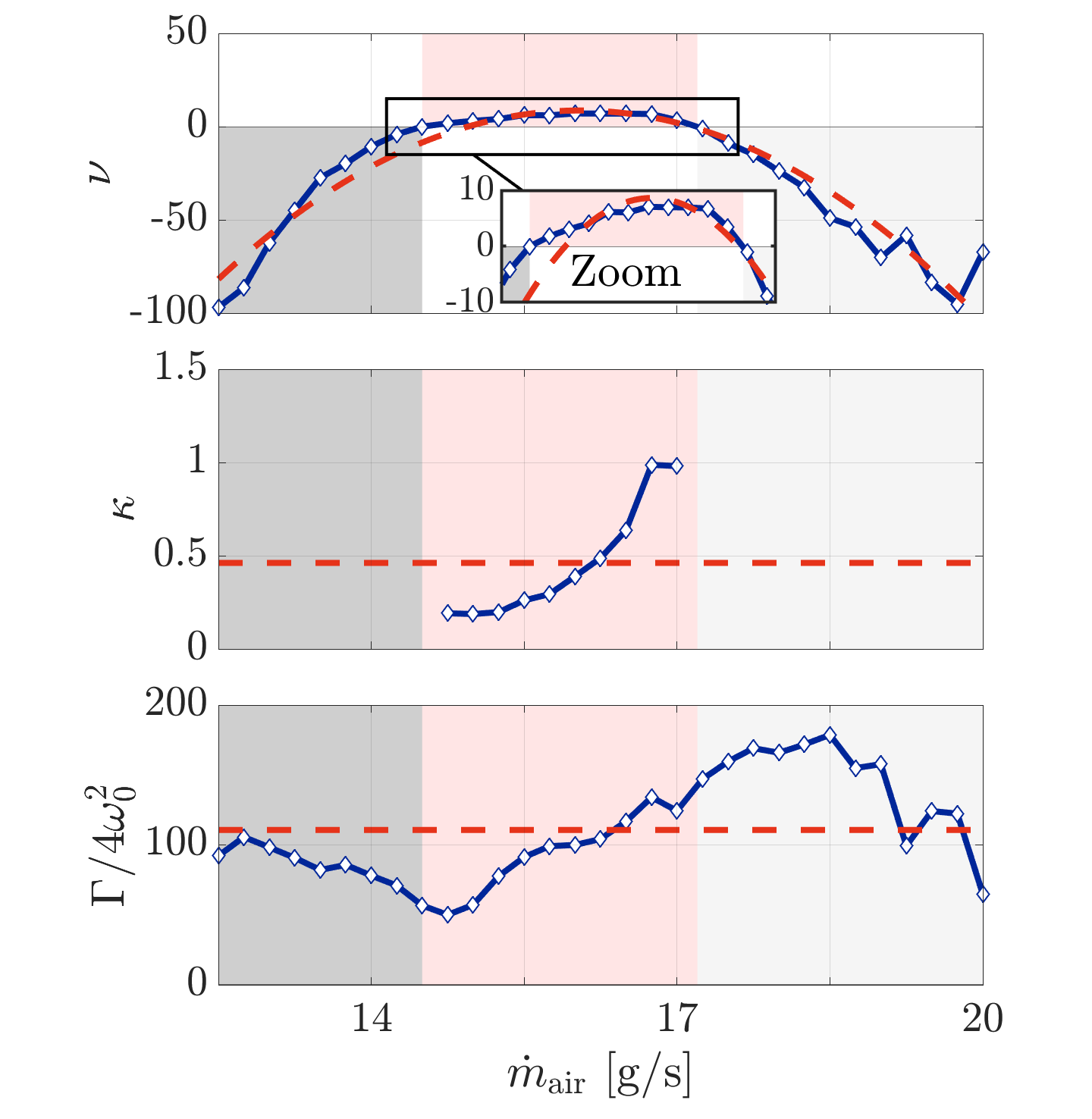}
    \caption{Identification of the low-order model parameters. The linear growth rate $\nu$, the cubic saturation constant $\kappa$ and the noise intensity $\Gamma$ are identified with the output-only method proposed in \cite{boujo2017robust}. The changes in the background colors of these three panels indicate a change in the linear stability of the system, as deduced from the sign of $\nu$. The dashed lines indicate the values adopted for $\nu(\dot{m}_\text{air})$, $\kappa$, $\Gamma$, in the simulations performed in section \ref{ss:ramp_sim}.}
    \label{fig:06_ID}
\end{figure}

Before scrutinizing the transitory dynamics, we introduce a low-order model of the thermoacoustic system, which  will be used to phenomenologically describe the observed physics. The thermoacoustic dynamics of the combustor is characterized by two consecutive and mirrored supercritical Hopf bifurcations, happening when the air mass flow rate is increased. An established model to describe thermoacoustic instabilities is the stochastically forced Van der Pol oscillator (VDP) \cite{li2013lock,terhaar2016suppression,noiray2017method}. In the present case, the linear damping coefficient is negative for low $\dot{m}_\text{air}$, then positive and finally negative again for high $\dot{m}_\text{air}$. The stochastic VDP dynamics is governed by:
\begin{equation}\label{eq:vdp}
    \ddot{p}+\omega_0^2p=[2\nu-\kappa p^2]\dot{p} + \xi(t),
\end{equation}
where $p$ designates the acoustic pressure in the combustor, $\omega_0$ the natural angular frequency (in the present case the one of the dominant eigenmode $\psi_0$), $\nu$ the linear growth rate, whose sign  defines the linear stability of the system, $\kappa$ a positive constant setting out the saturation of the thermoacoustic oscillations to a limit cycle, $\xi(t)$ a gaussian white noise of intensity $\Gamma$ that represents the stochastic forcing exerted by turbulence \cite{bonciolini2017output}. Given the fact that thermoacoustic instabilities in combustion chambers are usually characterized by $|\nu|\ll\omega_0$, there are two relevant time scales to the present problem, the fast time scale of the acoustic oscillation $T_0=2\pi/\omega_0$ and the slow one $T_\nu=2\pi/\nu$ associated with the amplitude dynamics, and one can therefore assume $p(t)=A(t)\cos{(\omega_0t+\varphi(t))}$.
For each operating point, and even for the ones that corresponds to thermoacoustic limit cycles, it is possible to identify the parameters of equation \eqref{eq:vdp} from acoustic pressure records, using the output-only identification methods presented in \cite{noiray2013deterministic}. Figure \ref{fig:06_ID} presents the result of the identification process, performed with the robust identification algorithm proposed in \cite{boujo2017robust}. Note that standard approaches based on a linear description of the problem, like for instance damping rate identification from estimates of the quality factor of the thermoacoustic peak, are inapplicable to operating regimes that corresponds to thermoacoustic limit cycles, where the nonlinearity of the flame response to high-amplitude acoustic perturbations plays a crucial role. \\
The non-monotonic dependence of the acoustic amplitude on $\dot{m}_\text{air}$ presented in fig.~\ref{fig:04_bifurcation}  corresponds to a non-monotonic evolution of the identified linear growth rate $\nu$. At low and high $\dot{m}_\text{air}$, this parameter is negative, and therefore the system is linearly stable, while for intermediate values of $\dot{m}_\text{air}$, $\nu$ is positive and the system exhibits a limit cycle. In the following sections, the dependence of linear growth rate on the air mass flow rate is approximated by
\begin{equation}\label{eq:nucos}
    \nu(\dot{m}_\text{air})=n_1\cos(n_2\dot{m}_\text{air})+n_3,
\end{equation}
 while $\kappa$ and $\Gamma$ are considered as constant, although they display a significant dependence on $\dot{m}_\text{air}$, by fixing their value to their mean over the considered range of $\dot{m}_\text{air}$ (red dashed lines in \ref{fig:06_ID}). This simplification is made because accounting for their dependence on $\dot{m}_\text{air}$ does not significantly impact the transient dynamics of this phenomenological model, as shown in appendix A.
%
%
\section{Ramp and Bifurcation Dodge}
\begin{figure}
    \centering
    \includegraphics[width=0.5\columnwidth]{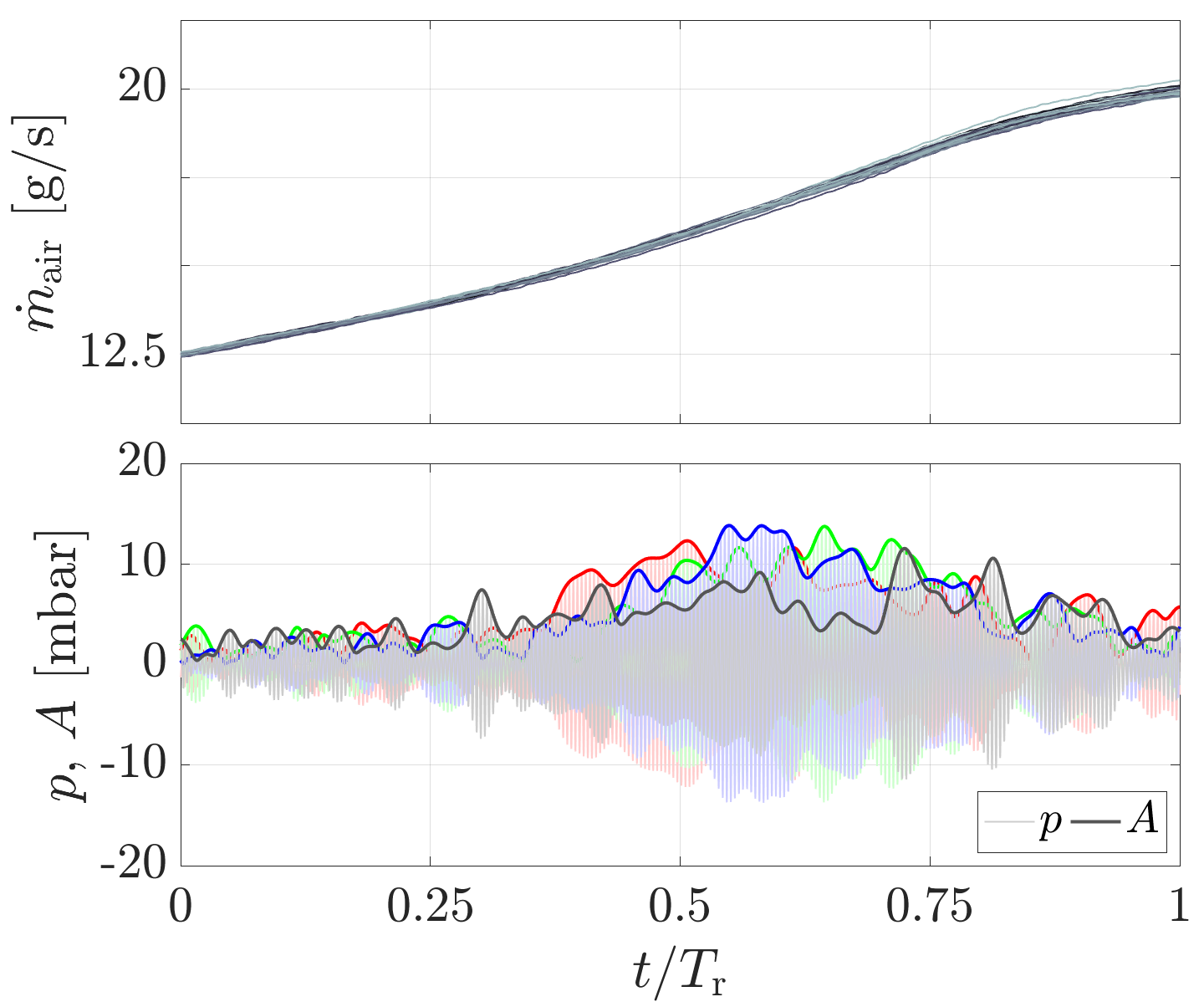}
    \caption{Result of one ramp experiment. Top: mass flow controller signal, showing the 50 ramps of $\dot{m}_\text{air}$. Bottom: four examples of the acquired  acoustic pressure $p$ and amplitude $A$ signals.}
    \label{fig:07_rampTT}
\end{figure}
\begin{figure*}
    \centering
    \includegraphics[width=\textwidth]{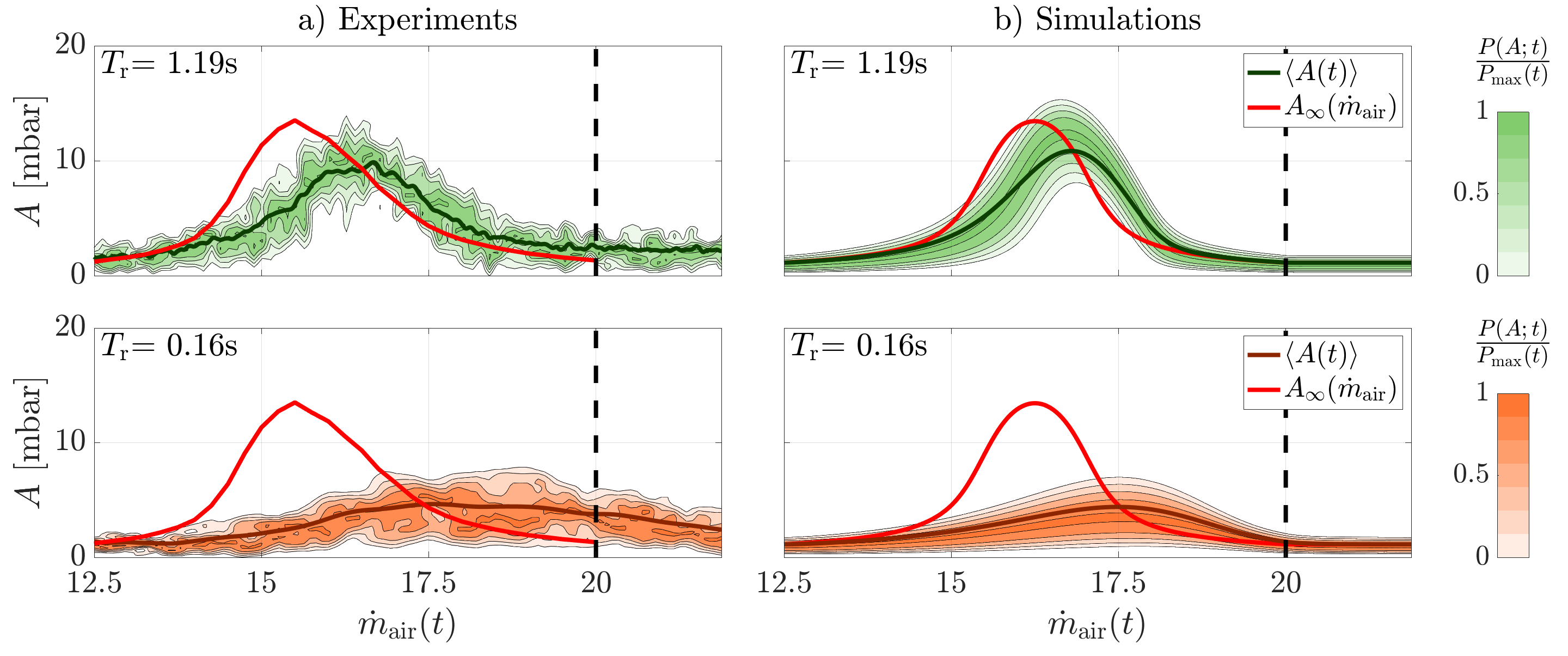}
    \caption{Amplitude PDF evolution under transient operation. a)  Experiments performed ramping $\dot{m}_\text{air}$. b)  Solution of the unsteady FPE \eqref{eq:fp}, with $\nu$ varying according to the cosine dependence on the control  parameter (eq. \eqref{eq:nucos}), which was identified from the experiments (see fig. \ref{fig:06_ID}). In both cases, the ramp is performed in the two indicated ramps time $T_\text{r}$. The red lines represent the reference stationary mean amplitude $A_\infty$.}
    \label{fig:07_ramp}
\end{figure*}
In this section, the dynamics of the system under transient operation is presented. As a preamble, it is important to mention that the present problem completely differs from the transient passage through resonances, obtained by sweeping the frequency of the external forcing of a linearly stable oscillator. The latter problem has been the subject of many studies in literature, and it is well known that the ramp rate of the harmonic forcing frequency has a substantial influence on the maximum amplitude reached during the transient \cite{white1971evaluation,irretier2000transient}. This is for example highly relevant for rotating machines at the startup \cite{chasalevris2012journal,fradkov2016control} or for hoist systems \cite{kaczmarczyk1997passage}.\\
In the present work, the nonlinear system exhibits self-sustained oscillations at fixed frequency $\approx\omega_0/2\pi$ for the middle range of the bifurcation parameter, and damped oscillations at about the same frequency for the lower and upper ranges of the bifurcation parameter. The transitions occur through two consecutive and mirrored Hopf bifurcations.
In the specific context of thermoacoustic instabilities in combustion chambers, one can note that thermoacoustic dynamics under operating condition transients is a topic that has received only very limited attention  \cite{culler2018effect,bonciolini2018effect,bonciolini2018experiments}, although real machines can be subject to rapid changes of operating conditions, like for instance aero-engines at take-off.\\
\subsection{Experimental results}
To perform the bifurcation parameter ramp test, an automatic controller was set to repeat 50 times a progressive opening of the air valve, such that $\dot{m}_\text{air}$ increased from 12.5 to 20 g/s in a quasi-linear fashion during each ramp.  The ramps duration is $T_\text{r}$. This experimental procedure is illustrated in fig. \ref{fig:07_rampTT}, where the time traces of the 50 ramps of the air mass flow rate (top panel) and 4 examples of the corresponding acoustic pressure and amplitude evolution (bottom panel) are presented. The acoustic pressure envelope statistics for this transient change of the bifurcation parameter are presented in fig. \ref{fig:07_ramp}a. In these plot one can see the evolution in time of the acoustic amplitude PDF $P(A;\,t)$ (contour map), and of its mean $\langle A(t) \rangle$ (thick line).
For ramp durations longer than 5 seconds, the transient PDF overlaps with the stationary one: no significant inertial effects are observed and the bifurcation parameter variation can be considered as quasi-steady. In the first row of the figure, the statistic of a ramp lasting $T_\text{r}$=1.19 s is reported. One can notice how inertial effects start to play a role: the development of the instability is delayed and the maximum attained amplitude lowered (see the stationary mean amplitude $A_\infty(\dot{m}_\text{air})$ plotted for reference as a red line). In the second row, the result of a test with a fast ramp lasting $T_\text{r}$=0.16 s is plotted. In this case, the onset of instability is almost suppressed: the acoustic level stays low and the bifurcation is dodged.
\subsection{Simulations results}\label{ss:ramp_sim}
We reproduced the dynamics observed in the experiment with the low-order model presented in section \ref{ss:vdp}.\\
The control parameter is ramped linearly over time at a certain rate $R$: $\dot{m}_\text{air}(t)=\dot{m}_0+R\,t$. $R$ is chosen to cover the whole $\dot{m}_\text{air}$ span in a time $T_\text{r}$. As a result, the linear growth rate of the model, whose dependence on $\dot{m}_\text{air}$ was approximated with a cosine (see eq. \eqref{eq:nucos}), varies over time: $\nu(t)=\nu(\dot{m}_\text{air}(t))$.\\
To obtain the statistic of the process, one can simulate many realizations of \eqref{eq:vdp} with this time-varying $\nu$ (Monte Carlo method). A computationally cheaper method is to solve the Fokker-Planck equation (FPE)  associated with the stochastic process $A(t)$ and directly obtain the evolution in time of the PDF $P(A;\,t)$:
\begin{equation}
\label{eq:fp}
\frac{\partial P}{\partial t}=-\frac{\partial}{\partial A}[\mathcal{F}(A,t)P]+\frac{\Gamma}{4\omega_0^2}\frac{\partial^2 P}{\partial A^2},
\end{equation}
where $\mathcal{F}(A,t)=A[\nu(t)-(\kappa/8)A^{2}]+\Gamma/(4\omega_{0}^{2}A)$, with a time-varying $\nu(t)$. All the other parameters are set to the mean of their identified values, as indicated in fig. \ref{fig:06_ID}. This last simplification does not have a significant impact on the solution: if the FPE is solved with varying $\kappa$ and $\Gamma$ coefficient, the result is not qualitatively different, as shown in the Appendix A.
The initial condition for the FPE is the stationary PDF $P_\infty(A)$ at the initial state, which is a function of the system's parameters and of the initial growth rate, and which can be derived analytically\footnote{$P_\infty(A;\, t=0)=\mathcal{N}A\exp{[{4\omega_{0}^{2}}/{\Gamma}({\nu_0 A^{2}}/{2}-{\kappa A^{4}}/{32})]}$, where $\nu_0=\nu(t=0)$ and $\mathcal{N}$ is a normalization constant. See \cite{bonciolini2017output} for more details.}. 
The solutions of the FPE, for the same $T_\text{r}$ as the experimental counterparts, are presented in fig. \ref{fig:07_ramp}b. One can appreciate how these two simulations closely follow the experimental statistic with a bifurcation dodge for $T_\text{r}=0.16$ s.
\begin{figure*}
\centering
    \includegraphics[width=\textwidth]{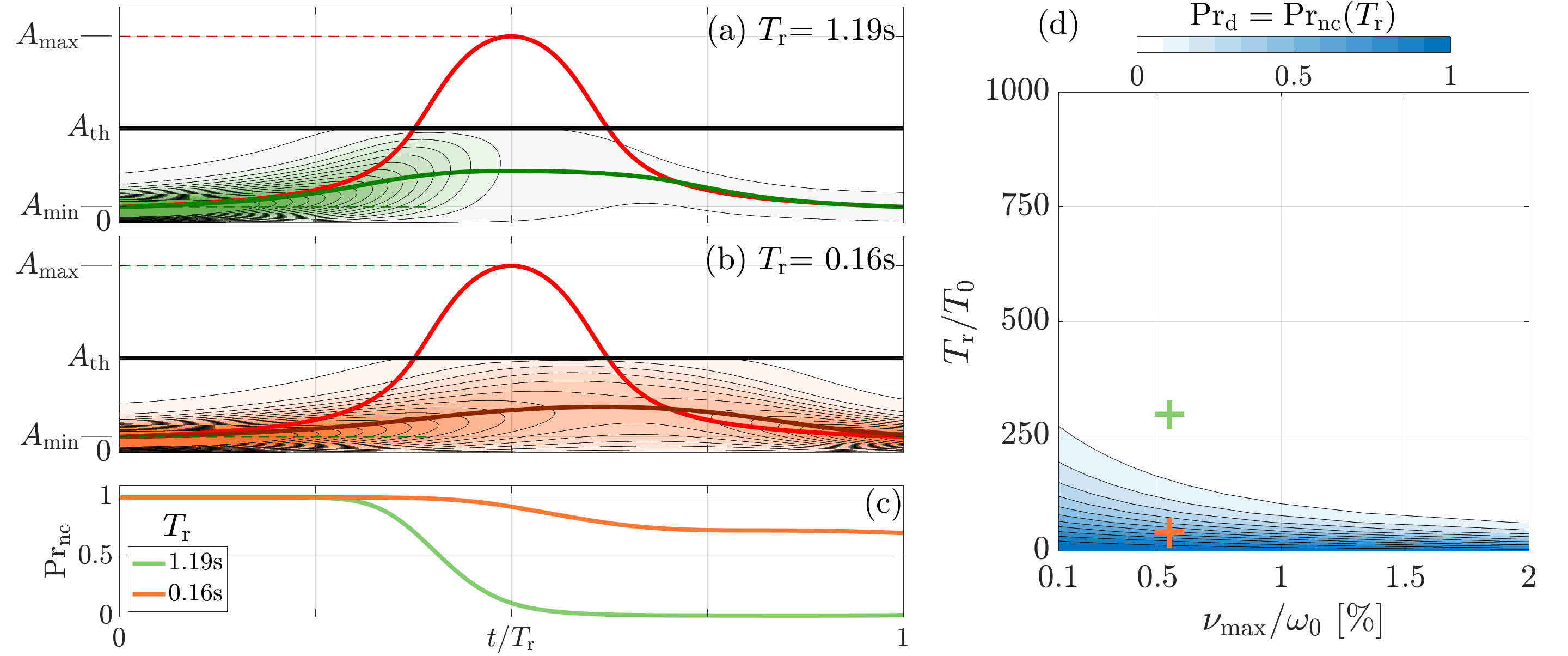}
    \caption{Calculation of the probability of dodging the bifurcation $\Pr_\text{d}$. Left: two examples of calculation, performed by solving the FPE with an absorbing boundary condition at $A=A_\text{th}$. The resulting PDFs $P(A;\,t)$ are shown in a) and b); the corresponding probabilities $\Pr_\text{nc}(t)$ of not having crossed $A_\text{th}$ before time $t$ are given in c). The values of $\nu_\text{max}=\nu(A_\text{max})$, with $A_\text{max}=\max\{A_\infty(\dot{m}_\text{air})\}$ and $T_\text{r}$ of these examples correspond to the ones of the experimental ramps presented in fig. \ref{fig:07_ramp}. d) summary map of the dodge probability $\Pr_\text{d}=P_\text{nc}(T_\text{r})$, as a function of $\nu_\text{max}$ and of $T_\text{r}$, respectively normalized with the oscillation angular frequency $\omega_0$ and period $T_0=2\pi/\omega_0$. The two crosses are the points presented in the left panels.}
    \label{fig:08_ProbNC}
\end{figure*}\\
The possibility of dodging the bifurcation is determined by the competition between the ramp time $T_\text{r}$ and the  characteristic time of the oscillation growth $T_\nu=2\pi/\nu$. When $T_\text{r}<T_\nu$, the range of linearly unstable conditions is overcome before the amplitude of the self-sustained oscillations can substantially grow, and a bifurcation dodge is possible. However, it is important to notice that the ramps considered in the present work nearly double the air mass flow, and they have durations $T_\text{r}$ that are very short compared to the ones of air and fuel mass flows variations of practical gas turbines or aero-engines. On the contrary, the present linear growth rates, and therefore the characteristic time $T_\nu$, are of the order as the ones that can be found in practice. Therefore, one can conclude  that dodging linearly unstable regions may be unrealistic in practice.\\
%
\section{Risk analysis}
In this section two analyses are performed using the phenomenological model given in eq. \eqref{eq:vdp}. The first aim is to quantify the probability of dodging the bifurcation, for a given linear ramping rate $R=(\dot{m}_\text{air}(T_\text{r})-\dot{m}_\text{air}(0))/T_\text{r}$ of the bifurcation parameter $\dot{m}_\text{air}$, and for a given maximum linear growth rate $\nu_\text{max}$, with eq. \eqref{eq:nucos}  defining  $\nu(\dot{m}_\text{air}(t))$ and  $\nu_\text{max}=\nu(A_\text{max})$, where $A_\text{max}=\max\{A_\infty(\dot{m}_\text{air})\}$. \\
The second aim is to estimate the  mean acoustic energy released when the level exceeds an amplitude threshold $A_\mathrm{th}$ during the transient. In some practical situations, this threshold could represent an amplitude of the output over which the system may be damaged.\\
It is assumed that this threshold is known a priori. Moreover, one consider cases for which the stationary behaviour, i.e. the evolution of the mean amplitude  as function of the control parameter, is also known. This amplitude is comprised between $A_\text{min}$ and $A_\text{max}$. However, the parameters $\nu$ and $\kappa$ are assumed to be unknown. These two parameters define the mean peak amplitude $A_\text{max}\approx\sqrt{8\nu_\text{max}/\kappa_\text{max}}$ (the exact analytic expression for the mean of $P(A)$ at any condition is given in Appendix B). 
For some practical applications, it is possible to guess a range of values for the linear growth rate $\nu$. For instance, in thermoacoustics $\nu$ generally varies between 0.1 and 2\% of the acoustic angular frequency $\omega_0$ \cite{ghirardo2018effect}. This analysis is thus performed by varying $\nu_\text{max}$ in that range, and adjusting $\kappa_\text{max}$ accordingly to maintain $A_\text{max}$ constant. Regarding the ramp duration $T_\text{r}$, it was considered the range going from 0 (step change of the control parameter) to 1000$T_0$, where $T_0=2\pi/\omega_0$ is the oscillation period.
%
\subsection{Bifurcation dodge probability}
With this test, the probability of not crossing a given threshold, or in other words the probability of dodging the bifurcation is computed. This is essentially an escape/first-passage problem, which is often addressed in random processes studies (e.g in \cite{gendelman2018escape,miller2012escape,hu2010first,ritchie2017probability,berglund2002beyond,cao2018linear,bishop2010stochastic}). This problem can be tackled in the current framework by solving the FPE with an absorbing boundary condition at the threshold amplitude $A_\mathrm{th}$, as described in \cite{bonciolini2018experiments}. In essence, during the bifurcation parameter sweep,  the amplitude PDF drifts and diffuses, and the   boundary condition at the threshold amplitude absorbs a fraction of the PDF. At every instant, what is left in the domain is the probability $\Pr_\text{nc}(t)=\int_0^{A_\text{th}} P(A;\,t)\text{d}A$ of not having crossed the boundary before that time $t$. The dodge probability can be defined as the probability of not having crossed the threshold $A_\text{th}$ before the end of the ramp $\Pr_\text{d}=\Pr_\text{nc}(T_\text{r})$. The results of this test are presented in fig. \ref{fig:08_ProbNC}. The diagrams a) and b) show the result of the FPE simulation for two combinations of ramp time $T_\text{r}$ with the same maximum linear growth rate $\nu_\text{max}$ (these points correspond to the experimental ramps presented in fig. \ref{fig:07_ramp}). The black line at the threshold amplitude $A_\text{th}$ is the absorbing boundary. 
In c), one can see how the probability of not having crossed $A_\text{th}$ decreases monotonically, with a steeper slope when the PDF drifts toward the boundary. 
In the fast ramp case (orange), the system state is only partially hitting the absorbing boundary and as a result the dodge probability is higher. On the contrary, for the slow case (green) the system follows more closely the stationary path, almost all the state probability is absorbed at the threshold level and the dodge probability is close to zero. The results of these simulations are reported in fig.~\ref{fig:08_ProbNC}d (green and orange crosses). This map represents the dodge probability $\Pr_\text{d}$, for different combination of $\nu_\text{max}$ and $T_\text{r}$, and for the same threshold amplitude $A_\text{th}$. 
\begin{figure*}
    \centering
    \includegraphics[width=\textwidth]{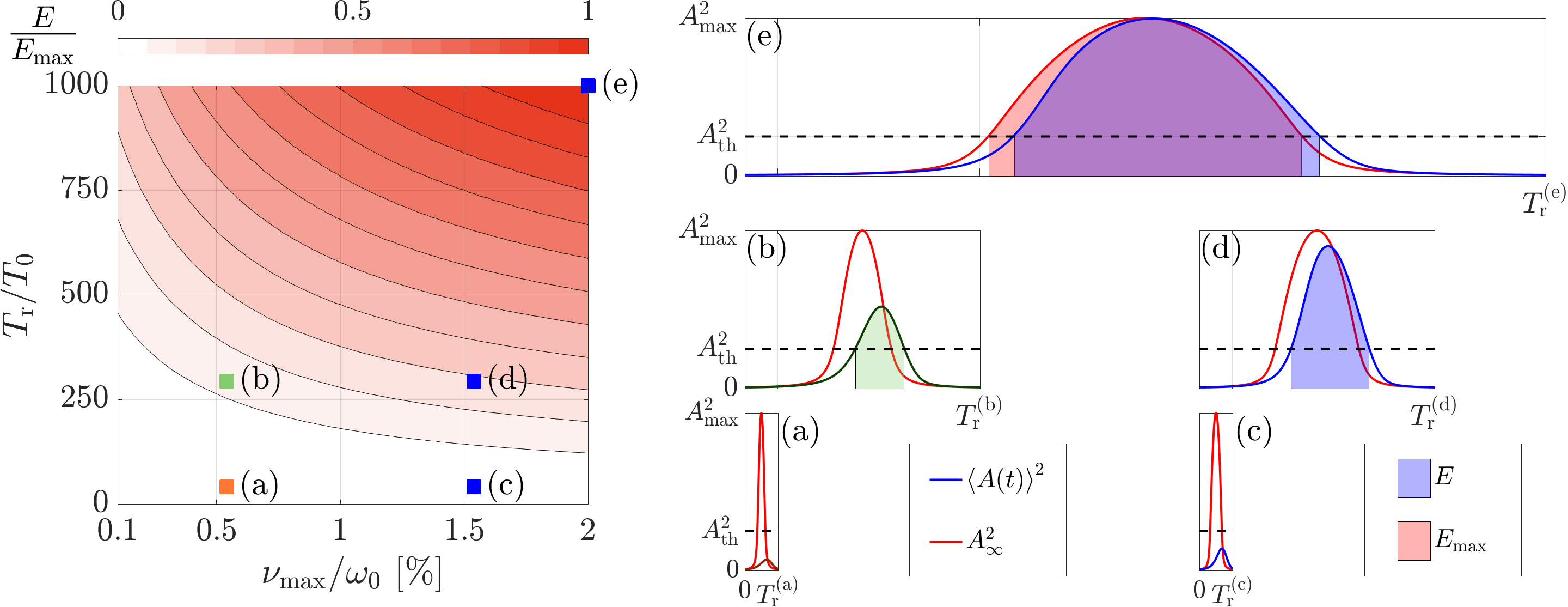}
    \caption{Cumulated energy released during the transient when the amplitude exceeds the threshold amplitude $A_\text{th}$, as defined in \eqref{eq:energy}, as a function of the parameters $\nu_\text{max}$ and $T_\text{r}$. The values of $E$ are normalized with $E_\text{max}=\int_{\Delta t_\text{th}}A_\infty^2(\dot{m}_\text{air}(t))\text{d}t$. In the right panels, the detail of the five points a--e indicated in the map. The red lines $A_\infty^2$, denote the reference  mean energy in a quasi-steady scenario. The blue curves ${\langle A(t)\rangle}^2$, are the mean amplitudes during the ramp. The areas highlighted in light blue/green below these curves, are the energies $E$. The area highlighted in light red in panel (e) is the reference energy $E_\text{max}$.}
    \label{fig:09_energy}
\end{figure*}
The linear growth rate at the initial state $\nu(0)=\nu_\text{min}$, the corresponding initial PDF and the mean amplitude $A_\text{min}$ are the same for all the points in the map and correspond to the ones identified from the experiments at $\dot{m}_\text{air}$=12.5 g/s.
As discussed earlier, $\kappa$ was varied according to $\nu_\text{max}$, in order to have the same mean peak amplitude $A_\text{max}$ for each simulation.
When $\nu_\text{max}$ is large, and therefore the characteristic oscillation growth time is short, a fast ramp is required to dodge even partially the bifurcation. 
The ramp time $T_\text{r}$ has a clear effect: decreasing it, which means performing shorter ramps with respect to the characteristic growth time, leads to higher dodge probability. The intuitive reasons behind this observation are that with shorter ramps i)  the characteristic diffusion time gets shorter than the characteristic drift time, and ii) the PDF drifts away from the stationary bifurcation path with a lower mean amplitude.
From this map, it becomes clear that being sure of dodging completely the bifurcation is difficult for realistic $\nu_\text{max}$ and for practical valve opening times. In other words, in practical systems exhibiting similar growth rates $\nu$, saturation constant $\kappa$ and stochastic forcing intensity $\Gamma$, it would be expected to cross the safety level, at least with a certain probability, even if the ramp is performed at a very high rate.\\

\subsection{Energy quantification}
Here, one computes the mean energy released by the self-sustained oscillator during the ramp and when the amplitude of the oscillations overcomes the threshold. This assessment is of primary importance, for instance for systems which are designed to resist to fatigue. What damages the system is the combination of the stress intensity and the number of cycles over the fatigue limit (which can be assimilated to our threshold amplitude $A_\text{th}$). Therefore we computed the mean over-threshold energy; during the ramp the mean amplitude  $\langle A(t) \rangle$ crosses two times the threshold. Therefore an over-threshold time interval $\Delta t_\text{th}$ can be defined. The mean energy released during this interval is defined as:
\begin{equation}\label{eq:energy}
    E = \int_{\Delta t_\text{th}}\langle A(t) \rangle ^2\,\mathrm{d}t.
\end{equation}
Figure \ref{fig:09_energy} is a map of this quantity as a function of $T_\text{r}$ and $\nu_\text{max}$. Five example points (a-e) are marked on the map and the corresponding transients are shown in the right panels. These panels show the evolution of ${\langle A(t) \rangle}^2$ compared to the stationary $A_\infty^2$, and the energy $E$ is highlighted (light blue area). The points (a) and (b) correspond to the experiments presented in the previous sections. 
The energy $E$ in the map is normalized with $E_\text{max}$, which is the mean energy that is released if the ramp time and the maximum linear growth rate are set at the extrema of their considered range (point (e) in the map), and if it is assumed that $\langle A(t)\rangle\equiv A_\infty(\dot{m}_\text{air}(t))$. This quantity is the red area in panel (e) on the right side of fig. \ref{fig:09_energy}, and it represents the ``worst case scenario'' energy that one could estimate from stationary data only.  
The value of $E$ is higher for higher $\nu_\text{max}$, since, as already discussed analyzing the previous test about the dodge probability, for the same ramp time the probability of dodging is lower and the maximum attained mean amplitude $\text{max}_t\{\langle A(t)\rangle\}$ is higher. Compare in this sense, the examples (b) and (d). For a certain $\nu_\text{max}$, the value of $E$ is higher for longer $T_\text{r}$, since the dodge probability decreases, the mean attained amplitude increases, and the time spent over the threshold is longer. For instance, comparing the two experimental points (a) and (b), it is clear how in the second case where the ramp time is longer, the system is able to cross, on average, the threshold and release energy. On the contrary, in case (a) $\langle A(t)\rangle$ stays below $A_\text{th}$ and the mean released energy is zero, even if that point has $\Pr_\text{d}<1$. This last example and all the other points in the white area of the map in fig. \ref{fig:09_energy} show that, even if a complete dodge of the bifurcation is not possible for a reasonable ramp time, a fast ramp guarantees to avoid, on average, the release of energy during the ramp, regardless of the (possibly) unknown value of the linear growth rate of the system.
\section{Conclusions}
In this paper, experimental evidences show that one can dodge consecutive and mirrored Hopf supercritical bifurcations, provided that the bifurcation parameter can be ramped fast enough. The required ramp rate is a function of the system's parameters, and in particular the ramp time $T_\text{r}$ needs to be shorter than the characteristic growth time $T_\nu= 2\pi/\nu_\text{max}$, where $\nu_\text{max}$ is the system's maximum linear growth rate in the bifurcation. A Van der Pol oscillator driven by additive white noise forcing is a sufficient model to reproduce the system dynamics in the transitory operation. The associated Fokker-Planck equation  is used to assess the probability of exceeding a certain oscillation amplitude threshold during the ramp, as a function of the system's parameters and of the ramp rate. This methodology can be applied to any system exhibiting such a sequence of mirrored bifurcations.
\section*{Acknowledgments}
This research is supported by the Swiss National Science Foundation under Grant 160579.

\section*{Compliance with ethical standards}

\section*{Conflict of interest}
The authors declare that they have no conflict
of interest.
\bibliographystyle{apsrev_mod}
\bibliography{./ghost.bib} 
\section*{Appendix A}
\begin{figure}
    \centering
    \includegraphics[width=0.5\columnwidth]{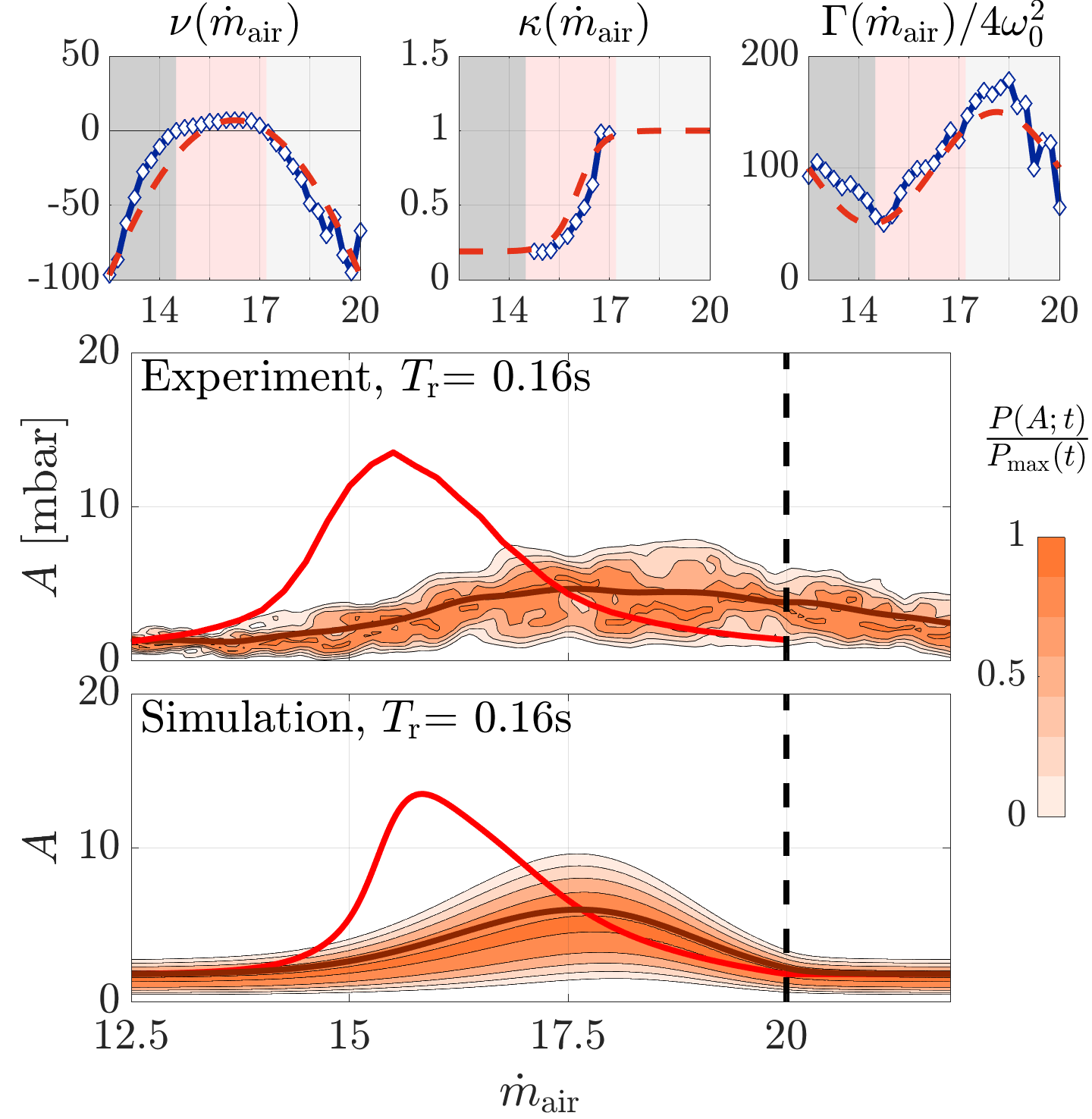}
    \caption{Simulation of the ramp dynamics with all the model's parameters varying. Top: approximation of the identified $\nu(\dot{m}_\text{air})$, $\kappa(\dot{m}_\text{air})$ and  $\Gamma(\dot{m}_\text{air})$ (blue lines), with simple functions (respectively a half cosine, a sigmoid and a cosine -- red lines). Center and bottom: Time evolution of the amplitude PDF $P(A;\,t)$ in the experiments and the one obtained solving the FPE with all the parameters varying with $\dot{m}_\text{air}$.} 
    \label{fig:10_full}
\end{figure}
Figure \ref{fig:10_full} shows the evolution of the PDF $P(A;\,t)$, when all the three parameters $\nu$, $\kappa$ and $\Gamma$ vary with the ramping of the bifurcation parameter, as shown in the top panels. In particular, the variation of the linear growth rate $\nu$ is approximated, as already done in the main body of the paper, with a half cosine fitted with the growth rates identified from the stationary experiments. The saturation constant dependence on the bifurcation parameter is approximated with the sigmoid function $\kappa(\dot{m}_\text{air})=k_1/(1+\exp{(-k_2\dot{m}_\text{air} +k_3)})+k_4$. The noise intensity variation is approximated with a cosine. One can observe in the bottom panel how the $P(A;\,t)$ obtained with this more complicated version of the model does not differ qualitatively from the one obtained with the simplified model and shown in fig. \ref{fig:07_ramp}, and reproduces qualitatively the experimental one, represented in the central panel for reference. It is important to remark that in this more complex model, the variation of each parameter is a direct function of the bifurcation parameter $\dot{m}_\text{air}$ only, and no inter-dependence between the three parameters $\nu$, $\kappa$ and $\Gamma$ is taken into account. That mutual dependence among parameters might be present in some physical system, including the one discussed in the present study. A further investigation of this aspect can be a topic for future investigations.
\section*{Appendix B}\label{sec:appendix}
Given the stationary PDF $P_\infty(A)$:
\begin{equation}\label{PAstat}
P_\infty(A)=\mathcal{N}A\exp\left[\frac{4\omega_0^2}{\Gamma}\left(\nu\frac{A^2}{2}-\kappa\frac{A^4}{32}\right)\right]
\end{equation}
with:
\begin{equation}
    \mathcal{N}=\bigg\{\sqrt{\frac{2\pi}{\kappa}\frac{\Gamma}{4\omega_0^2}}\exp\left(2\frac{\nu^2}{\kappa}\frac{4\omega_0^2}{\Gamma}\right) \left[1+ \text{erf}\left(\nu\sqrt{\frac{2}{\kappa}\frac{4\omega_0^2}{\Gamma}}\right)\right]\bigg\}^{-1}
\end{equation}
The mean of the distribution is given by:
\begin{equation}
\begin{split}
    \langle A \rangle ={} 
    & \mathcal{N}\pi\sqrt{\frac{2}{|\nu|\kappa^3}}\exp\left(\frac{\nu^2}{\kappa}\frac{4\omega_0^2}{\Gamma}\right) \bigg\{2\frac{\nu^3}{|\nu|}\left[I_{-\frac{1}{4}}\left(\frac{\nu^2}{\kappa}\frac{4\omega_0^2}{\Gamma}\right)+ 
        I_{\frac{3}{4}}\left(\frac{\nu^2}{\kappa}\frac{4\omega_0^2}{\Gamma}\right)\right] \ldots\\
    & \ldots+\left(\frac{\Gamma}{4\omega_0^2}\kappa+2\nu^2\right)I_{\frac{1}{4}}\left(\frac{\nu^2}{\kappa}\frac{4\omega_0^2}{\Gamma}\right)\ldots+2\nu^2I_{\frac{5}{4}}\left(\frac{\nu^2}{\kappa}\frac{4\omega_0^2}{\Gamma}\right)\bigg\},
\end{split}
\end{equation}
where $I_z(x)$ is the Modified Bessel function of first kind. This formula is indefinite if $\nu=0$, where it tends to:
\begin{equation}
    \langle A \rangle = \frac{2^{\frac{5}{4}}}{\sqrt{\pi}}\left(\frac{\Gamma}{4\omega_0^2\kappa}\right)^\frac{1}{4}\gamma\left(\frac{3}{4}\right),
\end{equation}
being $\gamma(x)$ the Euler gamma function.
\end{document}